\documentclass{article}

\usepackage{arxiv}

\usepackage[utf8]{inputenc} 
\usepackage[T1]{fontenc}    
\usepackage{hyperref}       
\usepackage{url}            
\usepackage{booktabs}       
\usepackage{amsfonts}       
\usepackage{nicefrac}       
\usepackage{microtype}      
\usepackage{lipsum}
\usepackage{graphicx}
\usepackage{cite}

\title{Attacks on Lightweight Hardware-Based Security Primitives}

\author{
  Jack~Edmonds \\
  Department of Electrical and Computer Engineering\\
  Santa Clara University\\
  Santa Clara, California, USA \\
  \texttt{jsedmonds@scu.edu} \\
   \And
 Fatemeh~Tehranipoor \\
  Department of Electrical and Computer Engineering\\
  Santa Clara University\\
  Santa Clara, California, USA \\
  \texttt{ftehranipoor@scu.edu} \\

}

\begin{document}
\maketitle

\begin{abstract}
In today’s digital age, the ease of data collection, transfer, and storage continues to shape modern society and the ways we interact with our world. The advantages are numerous, but there is also an increased risk of information unintentionally falling into the wrong hands. Finding methods of protecting sensitive information at the hardware level is of utmost importance, and in this paper we aim to provide a survey on recent developments in attacks on lightweight hardware-based security primitives (LHSPs) designed to do just that. Specifically, we provide an analysis of the attack resilience of these proposed LHSPs in an attempt bring awareness to any vulnerabilities that may exist. We do this in the hope that it will encourage the continued development of attack countermeasures as well as completely new methods of data protection in order to prevent the discussed methods of attack from remaining viable in the future. The types of LHSPs discussed include physical unclonable functions (PUFs) and true random number generators (TRNGs), with a primary emphasis placed on PUFs.
\end{abstract}


\section{Introduction and Motivation}
Perhaps the hottest topic in recent years regarding LHSPs is the discussion around physical unclonable functions (PUFs). The basic notion of a PUF involves a unique and reliable set of challenge-response pairs (CRPs), or inputs and outputs to a physical function, that can be used for security purposes including device identification, authentication~\cite{yan2015novelway}, and key generation. For a specific challenge, a PUF returns an associated response by way of a one-way function, meaning that the challenge cannot be re-obtained from the response. The size of the CRP set characterizes a PUF as being either strong (having a large enough number of CRPs that recording them all would be infeasible) or weak (having a small enough number of CRPs that recording them all would be feasible). Depending on which category the PUF falls into, it may or may not be suited for a specific use case. A number of PUF implementations exist, and a shared benefit of these LHSPs is that a PUF can provide device security without the need for additional hardware components intended solely for a security purpose. This is because PUFs take advantage of the inherent differences between silicon chips on the micro or nano scale that arise in the manufacturing process. They use these differences to provide a unique set of seemingly random yet reliable responses to corresponding challenges. This method of taking advantage of inherent uniqueness make PUFs especially desirable in situations where small size is a necessary constraints, such as in the case of Internet of Things
(IoT) technology~\cite{hassan2017internet}~\cite{tehranipoor2017invest}.

As discussed in~\cite{tehranipoor2017exploring}, employing secure methods of IoT device authentication and data protection is critical, yet poor security design is a common issue facing the IoT world~\cite{wortman2017proposing}, including devices on the consumer electronic market, in the medical and healthcare domains, and in other areas. In~\cite{tehranipoor2018low}, Tehranipoor et al. discuss consumer electronic device authentication and propose a low-cost PUF verification solution that uses keys generated from biometric signals~\cite{karimian2016evolving}. Another consumer electronics authentication scheme, termed P2M-Sec, is proposed in~\cite{wortman2018p2m} that uses PUF responses as seeds for a pseudorandom number generator. Multiple PUFs are investigated for use in the scheme, including an arbiter, ring oscillator~\cite{aguirre2020systematic}, bistable ring oscillator, butterfly, and DRAM-based PUF~\cite{commercial2017}. Similarly in~\cite{karimian2018secure}, Karimian et al. propose a PUF-based authentication framework with particular emphasis on ECG biometrics. The framework combines PUFs with hardware obfuscation, and makes non-volatile storage of keys (which can leave keys vulnerable to attack) unnecessary. The authors detail several case studies with their framework, and report that using the framework with an ECG provides the best trade-off between reliability, key length, entropy, and cost. In~\cite{karimian2017noise}, the authors discuss key generation techniques with ECGs based on different feature extraction techniques with varying noise sources. They propose the NA-IOMBA (noise aware interval optimized mapping bit allocation) protocol for improved key reliability, building off of their previously proposed IOMBA technique~\cite{karimian2016highly}. In~\cite{karimian2019heartid}, using ECG-based authentication is explored in the context of autonomous cars (HeartID), with heart signals being used to authenticate interaction between a person and a car (such as allowing entry to the vehicle).

\begin{figure}
    \centering
    \includegraphics[width=0.8\linewidth]{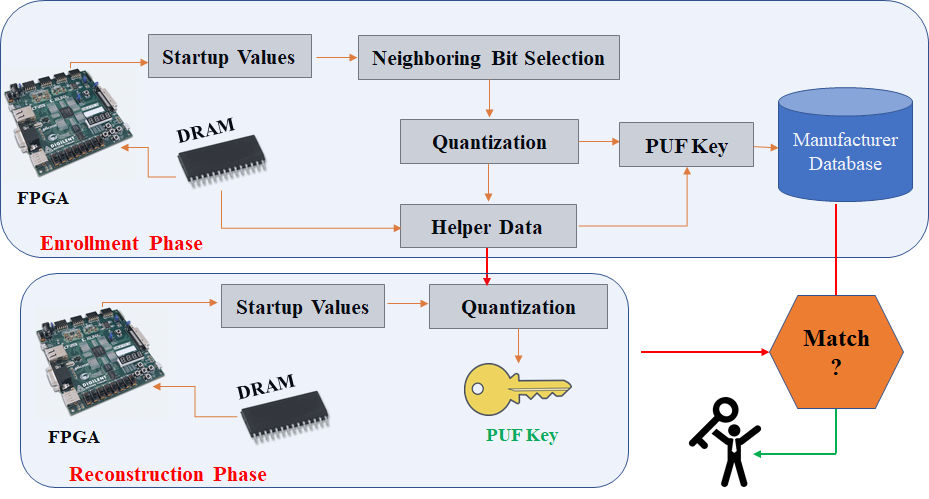}
    \caption{DRAM-based PUF key generation~\cite{karimian2019generate}.}
    \label{fig:noisydram}
\end{figure}

\begin{figure}[htp]
    \centering
    \includegraphics[width=0.8\linewidth]{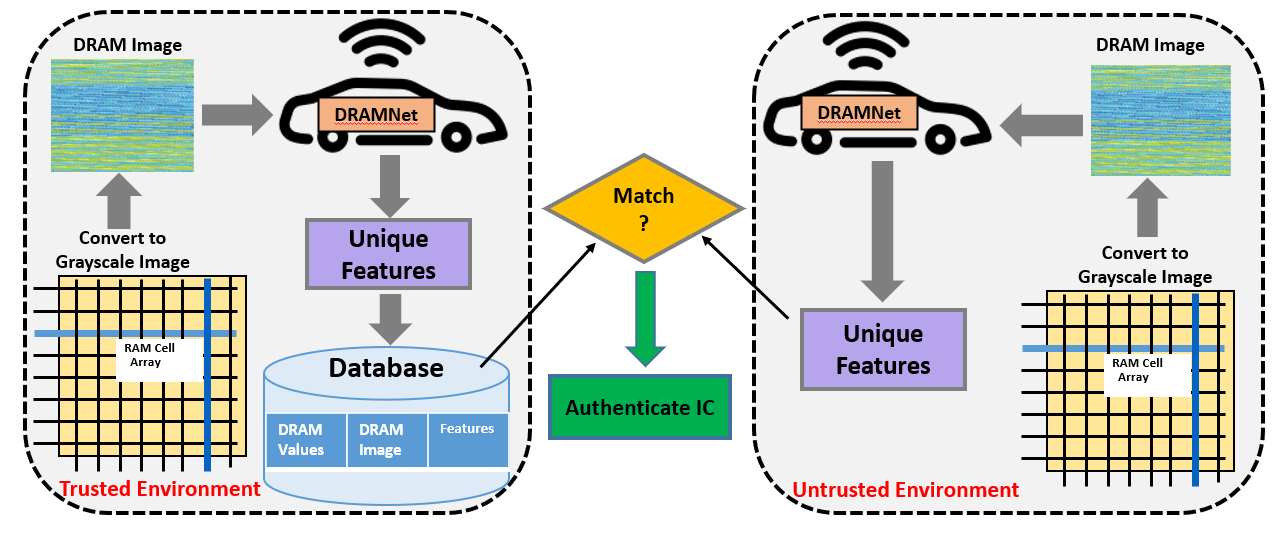}
    \caption{Authentication with DRAMNet~\cite{karimian2019dramnet}.}
    \label{fig:dramnet}
\end{figure}

One type of module often inherent in IoT devices that has been proposed for use as a PUF (as well as a TRNG) is dynamic random-access memory~\cite{anagnostopoulos2018overview}. Fig.~\ref{fig:noisydram} shows a key generation scheme proposed in~\cite{karimian2019generate} that can be used for device identification with DRAM-based PUFs, such as that proposed in~\cite{tehranipoor2016dram}~\cite{tehranipoor2015dram} by Tehranipoor et al. The 128-bit PUF relies on the random startup values of cell capacitors in an on-board DRAM module in order to generate a unique response when the device is turned on. In this case, the power-up sequence would be regarded as the challenge, and the values of the cells of interest make up the corresponding response. As seen in Fig.~\ref{fig:dramnet}, a similar use case of the entropic behavior observed in DRAM has been proposed in~\cite{karimian2019dramnet}~\cite{dramnet2} for an authentication scheme called DRAMNet. It uses a convolutional neural network (CNN) to extract and classify unique features from a DRAM image structure (created by converting the power-up sequence values into a 2-dimensional array) in order to authenticate a device with greater than 98\% accuracy and precision.

Several quality metrics are used to evaluate PUFs, including uniqueness, uniformity, and reliability. Uniqueness can be determined by calculating the inter-chip Hamming distance, which determines how different a PUF's responses will typically be when the same PUF is implemented on different chips and given the same challenges. Ideally, the inter-chip Hamming distance will be 50\%, meaning half of the bits in a PUF's response on one chip will be flipped when it's given the same challenge on another chip. Uniformity refers to the ratio of ones to zeros in PUF responses. Ideally, a PUF should exhibit 50\% ones and 50\% zeros on average in its responses in order to keep bit predictability as low as possible. Determining a PUF's reliability involves examining how well its CRP set holds up under varying environmental conditions. Ideally, a response will be exactly the same each time it's given the same challenge under all operating conditions, such as at different temperatures or after aging. In~\cite{anagnostopoulos2018securing}, DRAM-based PUFs, namely retention-based and row hammer PUFs, are evaluated in the context of temperature and voltage variations, and are shown to be highly dependent on temperature. The effect of temperature variation on PUFs is further investigated in~\cite{anagnostopoulos2018addressing}, and a method is proposed to account for the negative impacts. Specifically, temperature sensors, which are often present in DRAM modules, are used with a DRAM-based PUF in order to create an authentication protocol suitable for varying temperatures that restricts access to PUF responses when it's determined that the operation could be compromised by the environment's temperature.

Much research is put into ensuring PUF reliability. For example, in a paper by Tehranipoor et al.~\cite{tehranipoor2017investigation}, a report is given on the reliability of three DRAM-based PUFs aged 18 months. The authors show that the PUFs underwent only slight losses in their reliability over this period of time - the best of the three dropping from 88.9\% reliability to 87.5\%, and the worst from 89.7\% to 81.9\% over the period of accelerated aging. Another example of a PUF proposed specifically with reliability in mind is a phase calibrated ring oscillator PUF that can be implemented on a field-programmable gate array (FPGA)~\cite{yan2017phase}. The PUF is shown to have better response reliability than that of a traditional ring oscillator PUF, which relies on inherent wire and inverter delay to generate responses. Other PUF metrics including power requirements and footprint are also very important (especially in the realm of IoT), but uniqueness, uniformity, and reliability are among the most crucial in regards to a PUF's intended operation.

True random number generators (TRNGs) are another form of LHSP. Unlike a pseudo-random number generator (PRNG) that is ultimately deterministic in generating seemingly random outputs, a TRNG is non-deterministic, meaning it does not use an algorithm to generate outputs (which would reveal its secrets if known). Rather, a TRNG relies on a physical process to generate random bit streams. A few examples that a TRNG could utilize for this purpose include phase noise, thermal noise, jitter, random telegraph noise, and the photoelectric effect~\cite{tehranipoor2016robust}. TRNGs are used in a wide variety of security applications, with cryptographic algorithms often relying on them for key generation. As an example of a TRNG, one proposed by Tehranipoor et al.~\cite{tehranipoor2016robust} utilizes the remanence effect of DRAM (where some information remains in the memory cells after powering down) in order to generate random cryptographic keys. DRAM is also proposed for use as a TRNG in~\cite{eckert2017drng}, but the startup values are used instead. Another TRNG proposed in~\cite{tehranipoor2017study} is based on power supply noise, and based on results from a NIST statistical test suite, the authors conclude that the variations in voltage exhibit enough entropy that they can be used to generate random bits characteristic of a TRNG. Similarly in~\cite{tehranipoor2018dvft}, a dynamic feedback voltage tuning (DVFT) power supply noise-based TRNG is proposed. Six different power supplies are evaluated, and Gaussian distributions are observed, demonstrating non-cyclostationary behavior that can be utilized in the context of true random number generation~\cite{tehranipoor2018towards}.

\section{Literature Review}
While the concept of a PUF or TRNG may be nice, actually implementing one in practice that is fully resistant to attack is another story. In many cases regarding PUFs, these supposed "unclonable" functions have in fact been shown to be clonable through modeling. Equivalently, numerous successful attacks have been mounted against TRNGs. These attacks can come in various forms - many of which will be discussed in this section. In particular, the following literature review for PUFs is primarily focused on recent reports on PUF modeling attacks via machine learning (ML) and PUFs designed with ML-resistance in mind. The following subsection discusses recent attacks on TRNGs.

\subsection{Attacks on PUFs}
In this subsection, we provide a summary of some of the most recent developments in machine learning-based attacks on PUFs, as well as proposed ML-resistant PUFs. In all cases, the proposed/attacked PUFs are thought to meet at least a minimum set of requirements for uniqueness, uniformity, and reliability.

Beginning with a paper by Nguyen et al.~\cite{nguyen2019interpose}, the authors present the Interpose PUF and evaluate it against ML attacks. The Interpose PUF is a strong PUF consisting of two XOR arbiters, and the authors conclude that it is not vulnerable to logistic regression (LR) and covariance matrix adaptation evolution strategies (CMA-ES). In~\cite{wisiol2020splitting} however, it is shown that the Interpose PUF, which was thought to be more secure than existing XOR arbiter PUFs due to its resistance to LR and CMA-ES, is vulnerable to deep learning. The attack times recorded are comparable to those for other XOR arbiter PUFs, and the authors present results showing that they were able to achieve an attack success rate of at least 88\% on several variations of the Interpose PUF with a multilayer perceptron (MLP) neural network.

In~\cite{delvaux2019machine}, an ML-based side-channel impersonator attack is proposed and used to evaluate five different arbiter PUFs, specifically the PolyPUF by Konigsmark et al., the OB-PUF by Gao et al., the RPUF by Ye et al., the LHS-PUF by Idriss et al., and the PUF-FSM by Gao et al. All PUFs are successfully modeled, and it is noted that often the physical security of PUF protocols is not given enough attention in the design phase as opposed to mathematical security, which can result in unseen vulnerabilities.

In a paper by Wisiol et al., the authors present an attack on the Lightweight Secure PUF that focuses on attacking the input transformations (that result in local minima) that can be used to model the PUF. The Lightweight Secure PUF was previously thought to be significantly more resistant to ML than traditional XOR arbiter PUFs, with the input transformations being introduced for this purpose. The authors of this paper show that unfortunately these input transformations can actually be exploited to achieve successful modeling times comparable to those recorded when attacking classic XOR arbiter PUFs. LR and a correlation attack are used in the evaluation. The authors also propose a potential low-overhead countermeasure using fix-point-free permutations as an input transformation alternative that does not result in local minima. They find that this technique is able to significantly increase the PUF's resistance to their proposed attack.

In~\cite{laguduva2019machine} Laguduva et al. introduce a non-invasive architecture-independent ML attack targeting PUF-based IoT nodes. The three PUFs evaluated are of arbiter, XOR, and Lightweight architectures, and the attack involves the characterization of a given PUF via an optimal function in order to determine whether an LR, random forest (RF), or a neural network ML algorithm should be used to model it. With this attack strategy they find that they're able to achieve a modeling accuracy of ~95.3\%. The authors, however, also propose a countermeasure (that they call "discriminator") to their attack which is able distinguish cloned PUFs from the original for secure authentication from a cloud server with an accuracy of ~96\%.

In~\cite{wang2019adversarial}, a CRP mechanism is suggested as a countermeasure to ML modeling attacks. Traditionally, a response for a given challenge is fixed (assuming proper reliability), but the authors propose adding a feature to strong PUFs that is capable of inverting responses with a chosen probability in order to throw off an attacker's modeling scheme. LR and CMA-ES are used to test the effectiveness of the proposed mechanism with a 64-bit arbiter PUF, and the authors find that the PUF's resistance does indeed improve with the countermeasure active. The prediction accuracy by LR without any response inversion is ~99\%, and when an inversion probability of 50\% is active, the accuracy drops to ~55\%. Similarly for CMA-ES, the prediction accuracy drops from ~99\% without any inversion to ~78\% with a 50\% probability of inversion.

In~\cite{zhuang2019strong}, a subthreshold current array PUF is proposed that is resistant to ML attacks. The PUF consists of two transistor arrays that exhibit nonlinear behavior due to the nonlinearity of the I-V characteristics for the subthreshold region of a MOSFET. Support vector machine (SVM) and LR algorithms along with neural networks are used to evaluate the PUF, which is shown to outperform the arbiter and 3-XOR PUFs it's compared against in terms of attack resistance. In~\cite{mahmoodi2019chipsecure}, a reconfigurable flash-based PUF called ChipSecure is proposed that takes advantage of random I-V characteristics in flash memory, consisting of two layers of primitive blocks and a shift register. The authors attack their proposed design with an MLP and report that achieved prediction accuracy is close to an ideal 50\%.

In~\cite{kroeger2020effect}, the authors present results on the effect of PUF aging on ML attacks. Specifically, they report that ML models trained with a dataset of CRPs recorded at a certain PUF age lose their effectiveness when attempting to predict on an attack dataset of CRPs collected after the PUF has aged.

\begin{figure}
    \centering
    \includegraphics[width=0.7\linewidth]{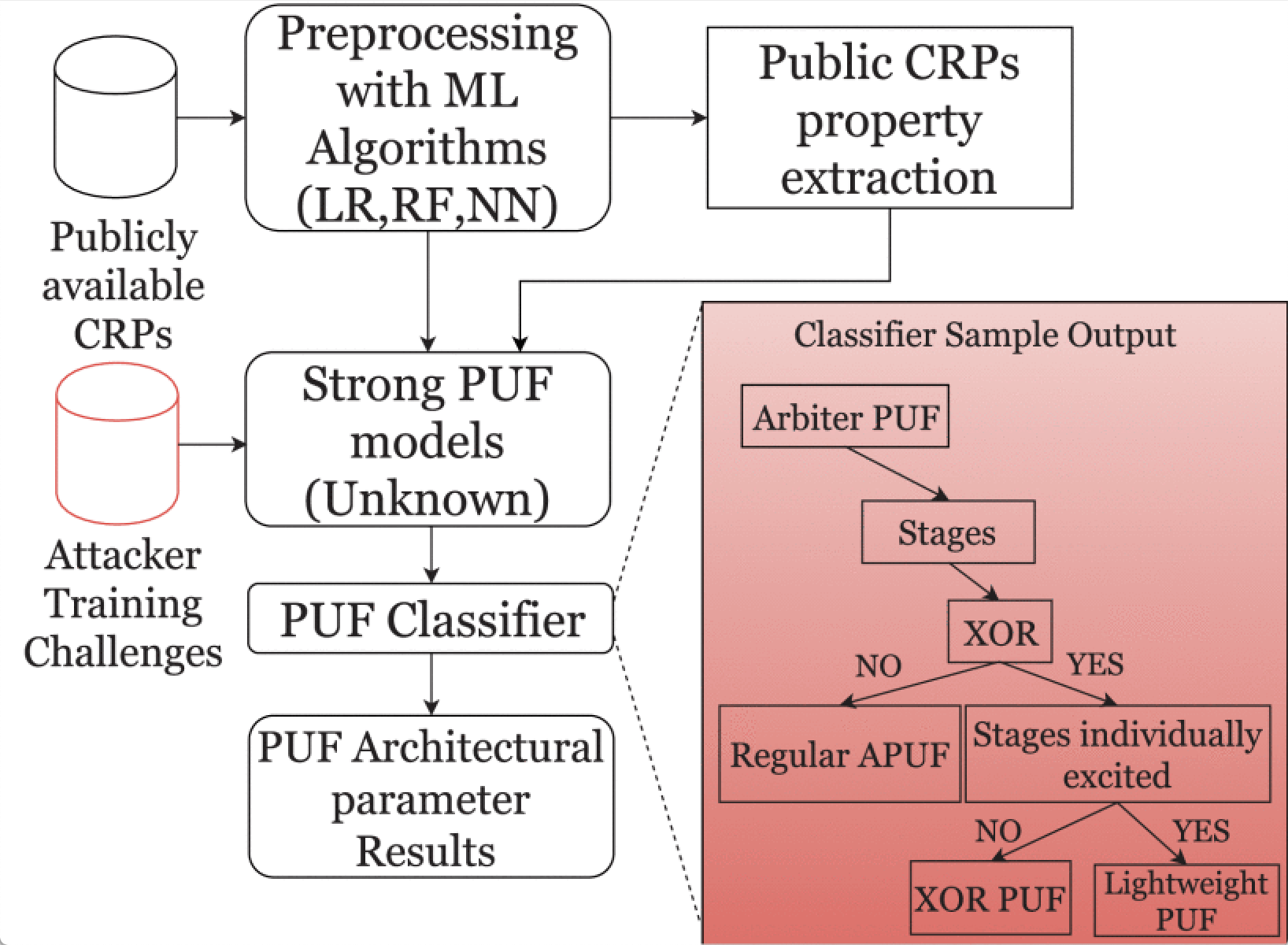}
    \caption{Architecture-independent PUF attack model~\cite{laguduva2019machine}.}
    \label{fig:archindep}
\end{figure}

In a paper by Wang et al., the authors proposed a Lattice strong PUF that's shown to be resistant to ML with both classical and quantum computing. The PUF is made with a physically obfuscated key, a learning-with-errors decryption function block, and a linear-feedback shift register (used as a pseudo-random number generator). They attack their PUF with SVM, LR, and deep learning neural networks, and find that the prediction accuracy is ~50\% after 1 million training CRPs in all cases. In~\cite{wu2020ct}, the authors propose a configurable tri-state strong PUF that combines an arbiter, ring oscillator, and bistable ring PUF for improved ML resistance over the configurable ring oscillator and dual-mode PUFs it's compared against. They attack all three PUFs with LR and neural networks, and report that while the dual-mode PUF's resistance to LR is similar to that of their tri-state PUF, they were able to achieve greater than 90\% modeling accuracy for the dual-mode PUF with neural networks, but only 50-60\% for their tri-state PUF. The configurable ring oscillator PUF is shown to be easily modeled by both LR and neural networks.

In a paper by Ge et al., the authors propose a challenge preprocessing structure for an arbiter PUF that's shown to improve ML resistance by hiding the linear correlation seen between a challenge and its corresponding response in a traditional arbiter PUF. Flip-flops are used to preprocess the original challenge before it's sent to the arbiter PUF, throwing off the direct relationship between the original challenge and the response that follows. The highest modeling accuracy achieved by the authors with linear regression, LR, SVM, and backpropogation neural network (BPNN) ML algorithms is ~61\% (with the BPNN).  In~\cite{pugazhenthi2019dla}, a DRAM-based PUF (DLA-PUF) that uses utilizes DRAM startup values is attacked with Naive Bayes (NB), LR, SVM, and CNN ML algorithms. It is shown to be resistant to these attack strategies when 3 DRAM startup values are used in its scheme.

In~\cite{ma2018machine}, the authors propose an arbiter-based multi-PUF that uses a group of weak PUFs to obfuscate challenges given to a strong PUF by taking the XOR of the weak PUF response bits and the challenge, and feeding the result into the strong PUF. The intention is to improve resistance to ML attacks. The authors attack their design with LR and CMA-ES, and find the prediction rate is ~50\% with LR (as opposed to 100\% when attacking a traditional arbiter PUF) and ~80\% with CMA-ES when the challenge size is at least 32 bits. In~\cite{wang2018machine}, a dual-mode PUF that works with both an odd number of inverters like a reconfigurable ring oscillator PUF and an even number of inverters like a bistable ring PUF is proposed, with the assumption that no ML attack can model both a ring oscillator and bistable ring PUF. Thus, as long as the current working mode can be hidden from attackers, its ML resistance should increase. They attack their PUF with LR and a neural network, and find that the LR accuracy is ~50\% and that the neural network accuracy is ~60\%, which is a significant improvement over the ML resistance of a ring oscillator PUF on its own (for which ~99\% modeling accuracy can be achieved).

In~\cite{xi2017strong}, the authors propose a subthreshold current array strong PUF that they show to be resistant to ML. The PUF boasts $2^{65}$ CRPs, and consists of two-dimensional transistor arrays with transistors operating in the subthreshold region with high voltage variability. The PUF is attacked with SVM, LR, and neural network algorithms, and from the authors' results is estimated to be 100 times more resistant to these attacks than a traditional arbiter PUF. In~\cite{tanaka2018coin}, a coin-flipping PUF is proposed that utilizes nonlinearity found in the convergence time of a bistable ring PUF with regard to threshold voltage variations. Responses are generated based on a ring oscillator's value at the convergence time of its associated bistable ring. The PUF is attacked with SVM, ES, RF, bagging, and boosting ML algorithms, and it is shown that the prediction accuracy is ~50\% across the board. In~\cite{pang2017novel}, the authors propose a double-layer 16Mb RRAM (resistive random-access memory) array PUF that uses the resistance values of cells to generate responses. The authors attack their double-layer PUF along with a single-layer version and a traditional arbiter PUF with a neural network, and are able to achieve a ~90\% success rate on the traditional arbiter, ~70\% on the single-layer version of their PUF, and only ~55\% on the double-layer version.

\subsection{Attacks on TRNGs}
This subsection discusses some of the most recent developments in TRNGs and their vulnerabilities. In a paper by Osuka et al., the authors investigate injecting sinusoidal waves through a power cable in order to non-invasively attack a TRNG by lowering its entropy and estimating the internal state of a ring oscillator through side-channel analysis. They show that when a signal at a particular frequency is injected, the TRNG's oscillator becomes locked, leading to a predictable generation of bits. The authors also propose a countermeasure to the attack, suggesting that including ferrite cores in the power line could help suppress the frequencies that are capable of locking the ring oscillator.

\begin{figure}
    \centering
    \includegraphics[width=0.7\linewidth]{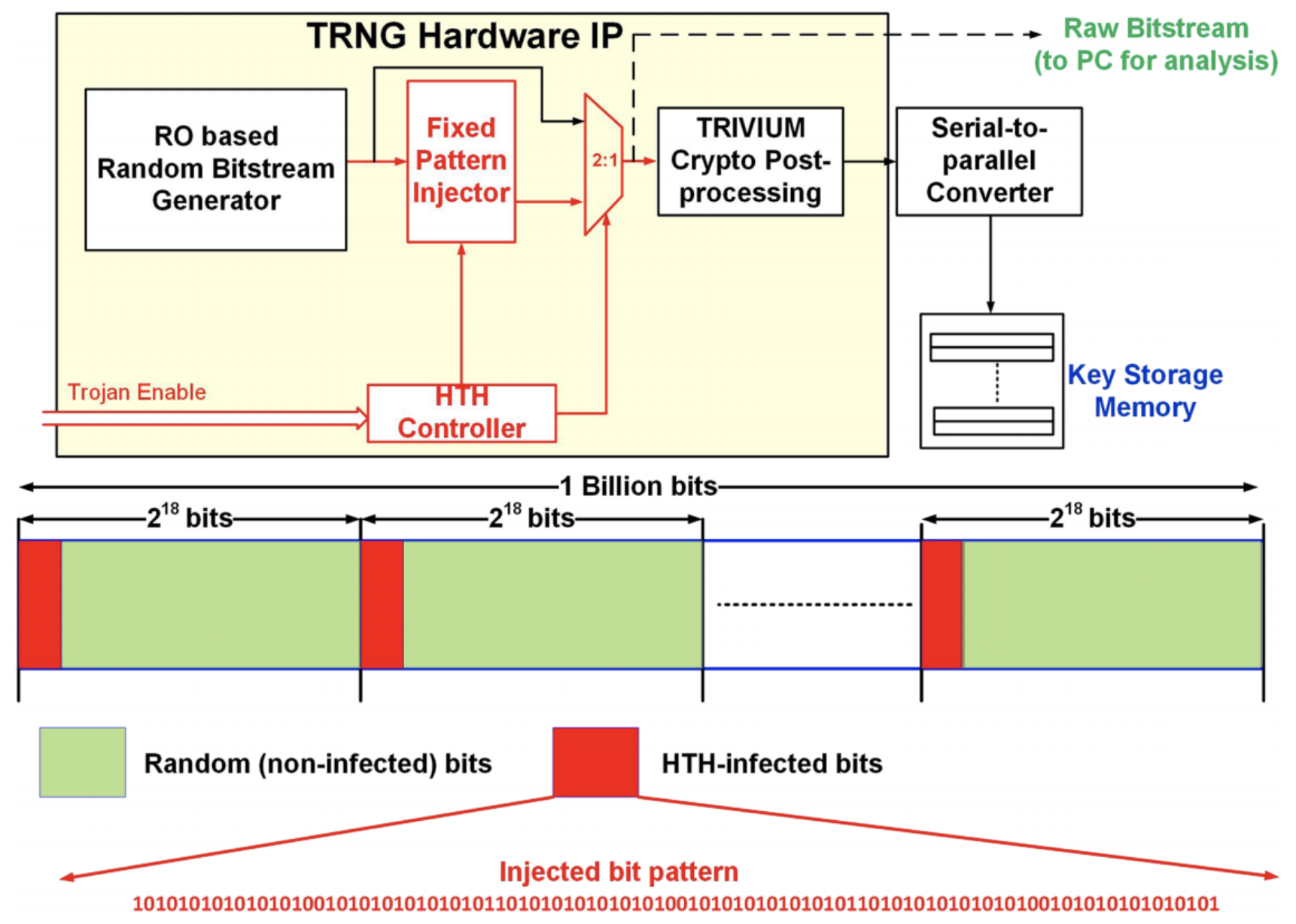}
    \caption{Hardware Trojan structure and impact~\cite{govindan2018hardware}.}
    \label{fig:tht}
\end{figure}

In a paper by Koyen et al., three forms of attack on a ring oscillator TRNG implemented on an FPGA are investigated. Of the three attack methods: voltage manipulation, ring oscillator locking, and replica observation, the voltage manipulation attack is reported to be the most effective. In a paper by Bonny et al., an attack on an FPGA-based non-autonomous chaotic oscillator TRNG is demonstrated. Their attack consists of applying clock glitches to the function clock in order to compromise the oscillator, and they show that with properly tuned attack parameters, they are able to disrupt the randomness of the generated bits and cause predictable behavior. In~\cite{govindan2018hardware}, a hardware Trojan-based attack on an FPGA-based TRNG is presented, as shown in Fig.~\ref{fig:tht}. The Trojan causes the TRNG to generate previously used keys with higher probability when active, while remaining undetectable otherwise. Specifically, the Trojan becomes active when the die temperature exceeds a set threshold. The authors use a transition effect ring oscillator-based Trojan detection technique along with a NIST statistical randomness testing suite to evaluate their proposed Trojan in terms of detectability, and find that both are unreliable and thus ineffective at detecting the Trojan.

Another hardware Trojan for ring oscillator-based TRNGs has been proposed by Ghandali et al. that generates predictable outputs via a stochastic Markov Chain model when triggered. The TRNG's source of entropy is also disabled at high temperatures, while exhibiting proper operation under normal operating conditions in order for the Trojan to go unnoticed. In~\cite{karimian2015genetic}, the authors propose a hardware Trojan detection scheme that uses ring oscillators to collect information from ICs in order to determine whether or not the IC is Trojan-free. The feature selection approach is based on a genetic algorithm that's combined with SVM, and is shown to have a ~30\% improvement in accuracy and ~97\% improvement in equal error rate over principle component analysis, which is one of the most commonly used methods of Trojan detection.

In a paper by Bahadur et al., the authors propose a side-channel resistant reconfigurable Galois ring oscillator-based TRNG implemented on an FPGA. Unlike traditional ring oscillator-based TRNGs, it is not dominated by any single frequency, and as a result is less vulnerable to frequency locking by electromagnetic and frequency injection attacks. In a paper by Japa et al., a Tunnel FET (field effect transistor) TRNG is proposed that relies on delays in an ambipolarity-based ring oscillator for its source of entropy. The authors deduce that their design is more resistant to reverse engineering attacks than traditional ring oscillator-based TRNGs because their design's FET transmission gates never switch off (which would cause the FETs to enter the tri-state region and stop oscillation), hindering an effective attack strategy for traditional ring oscillator-based TRNGs.

In~\cite{prada2020auto}, the authors propose a two-oscillator, two-counter auto-calibrated ring oscillator TRNG implemented on an FPGA that relies on accumulated oscillator jitter as its source of entropy as opposed to instantaneous jitter, which can make traditional ring oscillator TRNGs prone to power supply injection attacks. In~\cite{luo2020high}, a TRNG that combines a ring oscillator with a chaotic cellular automata topology is proposed. The design is tested through HSpice simulations, on an FPGA, and on an ASIC, and evaluated against three different attacks, namely frequency injection, power, and thermal attacks. From their experiments, the authors show that their TRNG is resistant to all three, suffering little entropy loss in all cases. In~\cite{yang2016total}, the authors propose an attack and TRNG failure detection scheme called TRNG On-the-fly Tests. The method is centered around monitoring the entropy of a TRNG after determining important statistical features, and they test their method on a traditional ring oscillator-based TRNG along with a carry-chain TRNG to show that their method is effective.

\section{Conclusions}
Maintaining awareness of recently discovered vulnerabilities in existing methods of data protection is vital for ensuring that we stay on top of what is secure, and what will leave information at risk. This encourages the continued development of improved security primitives and protocols, leading to more effective protection of users' data in scenarios where sensitive information is handled. We hope to have provided the reader with a brief yet informative overview of where PUF and TRNG LHSPs stand today, and in what direction the field of lightweight hardware-based security is moving~\cite{tehranipoor2017design}.


\end{document}